\documentclass[aps,prl,twocolumn,floatfix,showpacs,amssymb]{revtex4-1}
\usepackage[utf8]{inputenc}
\usepackage[english]{babel}
\usepackage{mathtools}
\usepackage{amsmath}
\usepackage{amsfonts}
\usepackage{amssymb}
\usepackage{graphicx}
\usepackage{float}
\usepackage{color}
\usepackage{braket}
\usepackage{mciteplus}
\def\urlprefix{  }
\def\url#1{ }
\newcommand{\be}{\begin{equation}}
\newcommand{\ee}{\end{equation}}
\newcommand{\ber}{\begin{eqnarray}}
\newcommand{\eer}{\end{eqnarray}}

\newcommand{\pv}{{\bf p}}

\newcommand{\Dv}{{\bf D}}

\newcommand{\Iv}{{\bf I}}

\newcommand{\qv}{{\bf q}}
\newcommand{\kv}{{\bf k}}

\newcommand{\vv}{{\bf v}}

\begin{document}

\title{Enhanced hydrodynamic transport in near magic angle twisted bilayer graphene}

\author{Mohammad Zarenia$^1$, Indra Yudishtira$^2$, Shaffique Adam$^{2,3,4}$, and Giovanni Vignale$^{1,2,3}$}
\affiliation{$^1$Department of Physics and Astronomy, University of Missouri, Columbia, Missouri 65211, USA\\
$^2$Yale-NUS College, 16 College Ave West, 138527 Singapore\\
$^3$Centre for Advanced 2D Materials, National
University of Singapore, 6 Science Drive 2, 117546, Singapore\\
$^4$Department of Physics, National
University of Singapore, 2 Science Drive 3, 117551, Singapore}

\begin{abstract}
Using the semiclassical quantum Boltzmann theory and employing the Dirac model with twist angle-dependent Fermi velocity we obtain results
for the electrical resistivity, the electronic thermal resistivity, the Seebeck coefficient, and the Wiedemann-Franz ratio  in near magic angle twisted bilayer graphene, as functions of doping density (around the charge-neutrality-point) and modified Fermi velocity $\tilde v$.  The $\tilde v$-dependence of the relevant scattering mechanisms, i.e. electron-hole Coulomb, long-ranged impurities, and acoustic gauge phonons, is considered in detail.  We find a {\it range} of twist angles and temperatures, where the combined effect of momentum-non-conserving collisions (long-ranged impurities and phonons) is minimal, opening a window for the observation of strong hydrodynamic transport.  Several  experimental signatures are identified, such as a sharp dependence of the electric resistivity on doping density and a large enhancement of the Wiedeman-Franz ratio and the Seebeck coefficient.   

\end{abstract}
\maketitle

\section{Introduction} 

Since the 2018 discovery of exotic superconductivity and correlated insulating phases in magic angle twisted bilayer graphene (tBLG) \cite{cao1,cao2,dean}, this system has been the subject of  many theoretical and experimental investigations, e.g. see Ref. \cite{ashvin} and references therein. The intense interest arises mainly from the new physics brought in by the low-lying flat bands near magic angles in tBLG \cite{macdonald}.	
The average Coulomb interaction~\cite{cea,rademaker} between the quasiparticles in narrow bands is far larger than the kinetic energy, giving access to  a strongly correlated regime and providing an ideal system for the observation of collective many-body phenomena. 

In this paper we focus on a particular collective phenomenon, namely  hydrodynamic transport in tBLG.
Hydrodynamic transport is expected whenever the momentum-conserving collisions between particles are much more frequent than the momentum-non-conserving collisions with impurities and/or lattice vibrations (phonons).  In addition, umklapp processes must be negligible.    Under these conditions the electric and thermal transport can be described by {\it hydrodynamic} equations for the flow of quasiparticles - electrons in the conduction band and holes in the valence band.  
Close to the charge neutrality point (CNP), where the densities of electrons and holes are nearly equal, a key indicator of the hydrodynamic regime is the ratio $\gamma=\tau_{\rm d}/\tau_{\rm{eh}}$ between the electron-hole scattering rate $1/\tau_{\rm {eh}}$ and the single-particle scattering rate $1/\tau_{\rm d}$ from momentum-non-conserving collisions with impurities {\it and} phonons. 
A large value of $\gamma \gg  1$ defines the so-called ``hydrodynamic transport window"~\cite{shaffique}, which has been theoretically predicted and experimentally observed in single-layer graphene~\cite{crossno,bandurin}, as well as in AB-stacked bilayer graphene~\cite{bandurin,morpurgo,zareniaB,wagner,tan}.

One important finding of the present work is that for some of the experimental samples in the literature e.g. Ref.~\cite{dean,polshyn}, these are already of sufficiently low disorder that our formalism predicts a robust hydrodynamic window close to twist angle of $1.11^\circ$. (From the very low temperature electrical transport, we estimate a charged impurity density of $1.6\times 10^{11}~\mathrm{cm}^{-2}$ for sample D5 in~\cite{dean}).  Other groups have data~\cite{cao1} where the impurity concentration is just above the threshold to observe hydrodynamic features, and therefore, the predictions we make below should be seen in cleaner samples in the near future.     

In the hydrodynamic regime, the electric and thermal transport have distinctive features that are described by the following expressions for the electric resistivity  $\rho_{\rm{el}}$ and the thermal resistivity $\rho_{\rm{th}}$, as functions of the dimensionless doping away from charge neutrality $\bar \mu\equiv\mu/(k_BT)$:
\be\label{eqRel}
\rho_{\rm{el}}(\bar\mu)\simeq \rho_{\rm C}\frac{\Gamma^2}{\Gamma^2+(\alpha\bar\mu)^2}\,, 
\ee
\be\label{eqRth}
\rho_{\rm{th}}(\bar\mu)\simeq  \rho_{\rm C}  \left[\Gamma^2+(\alpha\bar\mu)^2\right]\left(\frac{e^2}{k_B^2T}\right)\,,
\ee
where $\alpha=4\ln2/9\zeta(3)\approx 1/4$ is constant and $\bar \mu \ll 1$.  In these formulas $\rho_{\rm C} \propto1/(e^2\tau_{\rm{eh}})$ is the electric resistivity due to Coulomb electron-hole scattering at CNP \cite{kashuba, sachdev, zareniaG,zareniaB}, and $\Gamma=(k_B/e) \sqrt{T\rho_{\rm{th}}(0)/\rho_{\rm{C}}}$ where $\rho_{\rm{th}}(0)$ is the thermal resistivity at charge neutrality, which is due to momentum non-conserving collisions only, because the thermal current density coincides with the conserved momentum density at CNP.  Making use of the conventional Wiedemann-Franz law for noninteracting systems we can express $\Gamma$ more incisively as a ratio of disorder and interaction contributions to the electric resistivity, i.e.,  
 \be\label{eqG}
\Gamma \simeq 0.35\sqrt{\frac{\rho_{\rm el,d}(0)}{\rho_C}}\propto \frac{1}{\gamma^{1/2}}\,,
\ee
and $\rho_{\rm el,d}(0)$ is the non-interacting electric resistivity collisions.
Thus, we see that the cumulative effect of all types of  disorder, e.g., charged impurities, phonons, etc., is included in the single parameter  $\Gamma$  which becomes effectively a measure of  ``hydrodynamicity".

As shown in Ref.~\onlinecite{zareniaG}, the derivation of Eqs.~(\ref{eqRel}) and~(\ref{eqRth}) requires that the conditions $\Gamma^2\ll 1$ and  $\bar\mu\ll 1$ be satisfied.  These conditions define a temperature window for the observation of hydrodynamic effects: $\bar \mu\ll 1$ implies that the temperature is not too low, and $\Gamma \ll 1$ excludes high temperatures, where the phonon contribution to the resistivity would become very large.  In practice intermediate temperature in a 50 K -100 K range are most suitable.  Eqs.~(\ref{eqRel}) and~(\ref{eqRth}) provide us with explicit analytical expressions for the resistivities as functions of doping density, via the chemical potential.


Following the  theory outlined above,  the purpose of this paper is to understand how the electric and thermal resistivity, as well as the Wiedemann-Franz (WF) and the Seebeck coefficient,  behave as functions of the angle-dependent Fermi velocity in tBLG near charge neutrality.  While  employing a linear Dirac model to describe the the low-energy bands of tBLG,  we note that the twist angle acts as a new knob to vary the Fermi velocity and thus  the strength of interactions.
An incomplete theory, taking into account only electron scattering from long-range impurities would suggest hydrodynamic effects to gain strength at higher temperatures and as the magic angle is approached.  Careful consideration of the role of gauge phonons, which remain unscreened and contribute strongly to the resistivity, \cite{shaffique2}  reveals quite a different reality.  A strong hydrodynamic regime is found in the vicinity of the magic twist angle and at rather low temperature range  ($10\rm{K}\lesssim T\lesssim 50 \rm{K}$) compared  to single and bilayer graphene systems  \cite{zareniaG,zareniaB}.  This enhanced hydrodynamic is traced back to the strong suppression of electronic screening in tBGL near magic angle.   

Experimentally, the signature of the hydrodynamic regime should be clear and strong in the electrical resistivity, which is predicted to decrease sharply as a function of increasing doping density as nearly-free electrons become available for conduction (see Fig. 3a).  By contrast, the noninteracting electric resistivity is nearly independent of density, as seen by comparing Figs. 3a and 3b. Differently stated, Eq.~(\ref{eqRel}) predicts that the electric conductivity (inverse of the resistivity) grows as $\bar \mu^2/\Gamma^2$ as one moves away from CNP.  The positive curvature of the conductivity versus density is thus proportional to $1/\Gamma^2$ and provides a direct measure of ``hydrodynamicity". 
Another striking signature of hydrodynamics, although more challenging to observe experimentally,  would be the value of the Wiedemann-Franz ratio between the electric and thermal resistivity at CNP, which, from Eqs.~(\ref{eqRel}) and (\ref{eqRth})  is seen to be proportional to $1/\Gamma^2$.  The position of a maximum in the Seebeck coefficient, which is predicted to occur at $\alpha \bar \mu=\Gamma$ would be yet another signature. 
The behavior of the key parameter $\Gamma$ as a function of temperature and twist angle is summarized in Fig. 4, which we hope will be a valuable guide to experimentalists hunting for signatures of hydrodynamic transport in tBLG.  

%
%
  
This paper is organized as follows.
In Sec. II, we first evaluate  the angle-dependence of the screened intrinsic electric and thermal resistivitities. In Sec. III, we obtain the resistivities associated with the long-ranged charged impurity as well as the gauge phonons. 
Having the key ingredients, i.e. $\rho_C$ and $\Gamma$, in Section IV we calculate the electric resistivities and the Seebeck coefficient, and show that the WF ratio -- a direct indicator of the hydrodynamic regime -- is strongly enhanced at CNP as the magic angle is approached.   Sec. V presents our outlook and conclusions. 
\section{Intrinsic resistivity}
At low-densities around the CNP, the energy spectrum of tBLG, can be approximated by the Dirac model $
\epsilon_{\kv,\pm}=\pm\hbar\tilde{v}k $ with a twist angle-dependent Fermi velocity $\tilde v$,
\be
\tilde v=\frac{1-3\lambda^2}{1+6\lambda^2},~~\lambda=\frac{w}{\hbar v\Delta K},~~\Delta K\approx k_D\theta
\ee
where $\tilde v$ is in units of  the graphene Fermi velocity $v\simeq10^6$ m$/$s, $\theta$ is the twist angle, $w=110$ meV is the interlayer hopping,   and  $k_D=4\pi/3a$ with $a=0.246$ nm is the graphene lattice constant. 
Comparing with the tight-binding band structure, the authors in Ref. \cite{shaffique2}, show that the Dirac model is a valid model at densities $n\lesssim 8\times10^{10}$ cm$^{-2}$ and for twist angles $\theta\gtrsim1^\circ$. 

\begin{figure}
\centering
\includegraphics[width=10cm]{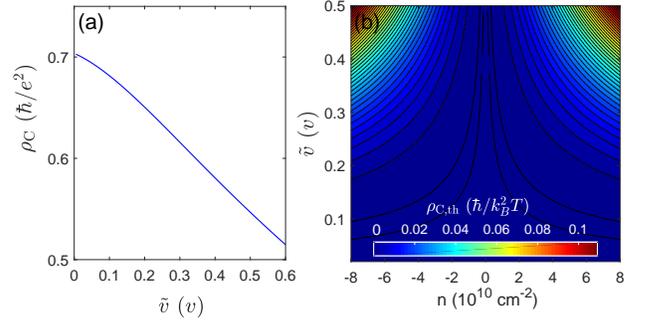}
\caption{(a) Electric resistivity $\rho_C$ from electron-hole Coulomb scattering at $\bar \mu=0$ (Coulomb drag resistivity) approaches a constant value as the modified Fermi velocity $\tilde v$ tends to zero.  $\rho_C$ also has a very weak (logarithmic) dependence on temperature, which is neglected here.   (b) 2D contour-plot of the intrinsic Coulomb thermal resistivity $\rho_{\rm{C,th}}\equiv \rho_C(\alpha\bar\mu)^2 e^2/(k_B^2T)$ obtained from Eq.~(\ref{eqRth})  in the limit of zero disorder ($\Gamma=0$) as a function of doping density $n$ and $\tilde v$ (in units of graphene Fermi velocity $v$) at fixed $T=50$ K.  Notice that $\rho_{C,th}$ vanishes at $\bar \mu=0$, reflecting the conservation of momentum.}
\label{fig1}
\end{figure}

In our recent work \cite{zareniaG} we have demonstrated that within a Dirac model and {\it in the absence of any disorder} (i.e., for $\Gamma=0$), the  electrical and thermal resistivities can be identified as
\be\label{CNPCoefficients}
\begin{split}
        \rho_{\rm{el,C}}(0)=\rho_{\rm{C}}=I_{\rm{C}} a^2/e^2,~\rho_{\rm{el,C}}(\bar \mu\neq0)=0,\\
        \rho_{\rm{th,C}}(0)=0,~~\rho_{\rm{th,C}}(\bar\mu\neq0)=I_{\rm{C}} b^2/(k_B^2T),
  \end{split}
\ee
where, $I_{\rm{C}}(\bar\mu,T)$ is the Coulomb collision kernel, given by Eq. (21) in Ref. \cite{zareniaG},  $a\sim \pi\beta\hbar^2/\ln 4$ and $b\sim[2\pi\beta\hbar^2/9\zeta(3)]\bar\mu$ for $\bar\mu\to0$. 
The {\it intrinsic} electric resistivity $\rho_{\rm{C}}$ (first calculated in Refs. \cite{kashuba, sachdev}  for graphene) is associated with the Coulomb drag between the electrons and holes at the CNP ($\mu=0$). In agreement with Refs. \cite{kashuba} and \cite{sachdev} we find that in the absence of screening
\be
\rho_{\rm{C}}^{\rm{~unscreened}}= I_{\rm{C}}^{\rm{~unscreened}}a^2 \propto \tilde v^{-2}.
\ee
With screening the bare Coulomb interaction $v_q$ is modified to, 
\be\label{EqVqv}
V(q,\omega\to0)=\frac{v_q}{|1-v_q\Pi_0|}
\ee
where the polarizability function $\Pi_0(\qv,T)$ is calculated numerically. However,  close to CNP and for  $T\gg T_F$,  the asymptotic form of  $\Pi_0(\qv,T)$ is given by \cite{revDasSarma,sarmaScreening},
\be\label{EqPi0T}
\Pi_0(\qv,T)=\frac{8k_F}{2\pi \hbar\tilde v}\left[\frac{T}{T_F}\ln4+\frac{1}{24}\frac{q^2}{k_F^2}\frac{T_F}{T}\right],
\ee
using which we can write, 
\be\label{eqRc}
\rho_{\rm{C}}=\frac{\rho_{\rm{C}}^{\rm{~unscreened}}}{|V(q,\omega\to0)|^2} \propto  (c+\tilde v)^{-2},
\ee
where $c$ has the dimension of a velocity and we obtain its  value numerically, e.g. $c\approx 1.2v$ at $T=50$ K.  In Fig. \ref{fig1}a we show the numerical results  for the screened $\rho_{C}$ as a function of $\tilde v$. In contrast to the $T$-dependent $\rho^{\rm{unscreened}}$, it is interesting to note that the screened $\rho_{C}^{\rm{screened}}$ has a very weak (logarithmic) dependence on temperature, which is neglected here \cite{sachdev}. At the magic twist angle ($\tilde v\to0$), we obtain  $\rho_{C}\approx 0.7~(\hbar/e^2)$. 

In Fig. \ref{fig1}b we have shown a 2D contour-plot of  the Coulomb thermal resistivity $\rho_{\rm{C,th}}$ as a function of doping density $n$ and Fermi velocity $\tilde v$ at fixed $T=50$ K. At the CNP,  $\rho_{\rm{C,th}}(0)$  vanishes for any value of $\tilde v$, see Eqs. (\ref{CNPCoefficients}).  Within the Dirac model
$n\approx 8\bar\mu/[\pi\beta^2 (\hbar\tilde v)^2]$ (note that the Dirac model breaks down at precisely magic twist angle $\tilde v=0$), and therefore away from the CNP  we obtain
\be
\rho_{\rm{C,th}}(\bar\mu\neq0)=\rho_{\rm{C}}(b^2/a^2)(e^2/k_B^2T)\propto \rho_{\rm{C}}~ \bar\mu^2 \propto \frac{\tilde v^4}{(c+\tilde v)^2}\,,
\ee
which exhibits a substantial suppression of $\rho_{\rm{C,th}}$ as $\tilde v \to 0$ as seen in Fig. \ref{fig1}b.
%
%
\section {Momentum non-conserving collisions}
\subsubsection{A. Long-range charged impurity}
\begin{figure}
\centering
\includegraphics[width=8.5cm]{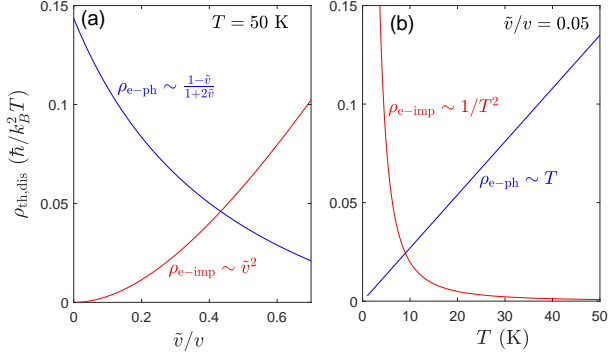}
\caption{Thermal resistivity associated with the long-ranged charge impurity (red curves) and acoustic gauge phonons (blue curves) as functions of (a) $\tilde v/v$ (at $T=50$ K) and (b) temperature $T$ (at $\tilde v/v=0.05$).  Both sets of curves are calculated at the charge-neutrality-point and the charged impurity density in (a) is set to $n_{\rm{imp}}=1\times10^{10}$ cm$^{-2}$. $\rho_{\rm{th,dis}}$ can be linked to the corresponding electric resistivity $\rho_{\rm{el,dis}}$, through the Wiedemann-Franz law for noninteracting electronic systems, which takes the form $\rho_{\rm{el,dis}}/\rho_{\rm{th,dis}}\simeq 2.4 \pi^2k_B^2T/(3e^2)$ for the Dirac model near charge neutrality.}
\label{fig2}
\end{figure}

We now evaluate the contribution of charged impurities and its $\tilde v$-dependence within the Dirac-modeled tBLG.  We recall that the  momentum non-conserving collision integral of the scattering potential of
randomly distributed screened (long-ranged) impurity charge centers is
given by \cite{revDasSarma},
\begin{equation}\label{eqd1}
\begin{split} 
&\Iv_{\rm{dis}}(\kv ,\eta)  = 8\times\frac{2\pi n_{\rm{imp}}}{\hbar}\sum_{\eta '}\sum_{\kv'}|\frac{v_q}{1-v_q\Pi_0(\qv,T)}|^2 F_{\kv ,\kv'}^{\eta\eta'}\times \\
&(f_{\kv ,\eta } -f_{\kv ',\eta '})\delta(\epsilon_{\kv ,\eta }-\epsilon_{\kv ',\eta '}),
\end{split}
\end{equation}
where $\eta (\eta')=\pm1$ sums over the two bands,  $\qv=\kv-\kv'$ and $n_{\rm{imp}}$ is the disorder density. The factor $8$ accounts for the spin and both graphene and moir\'{e} valley degeneracies. Inserting the non-equilibrium distribution function $f=f_0+f_0'\vv_\kv\cdot (\pv_n+\beta\tilde{\epsilon}_{\kv,\gamma}\pv_s)$, where $\pv_n$ and $\pv_s$ are the momentum shifts due to the charge and heat (entropy) currents respectively, we write the linearized collision kernels as
\be\label{EqIimp}
\begin{split}
&\rm{I}_{\rm{imp}}^{(m)} =\frac{4n_{\rm{d}}}{\hbar^3}\left(\frac{e^2}{\kappa}\right)^2\int_{0}^{2\pi}  d\theta~ (1+\cos\theta)\times \\ &\int_{0}^{\infty} d\bar\epsilon~ \frac{~(\bar\epsilon-\bar\mu)^m f'(\bar\epsilon) + (-1)^\alpha (\bar\epsilon+\bar\mu)^m f'(-\bar\epsilon)}{|\varepsilon(\bar\epsilon,\theta)|^2},
\end{split}
\ee
where $m=0,1,2$ give the matrix elements  $\rm{I}_{\rm{imp}}^{11}$, $\rm{I}_{\rm{imp}}^{12}=\rm{I}_{\rm{imp}}^{21}$,  $\rm{I}_{\rm{imp}}^{22}$, respectively.  $\bar\epsilon=\beta\epsilon_{\kv}$, $f'(\bar\epsilon)= \exp(\bar\epsilon-\bar\mu)/(1+\exp(\bar\epsilon-\bar\mu)^2$,  and using Eq. (\ref{EqPi0T}) the dielectric function $\varepsilon(\bar\epsilon,\theta)$ is
\be
\varepsilon(\bar\epsilon,\theta)=1+ \frac{8(e^2/\kappa)}{\hbar \tilde v~\sqrt{2(1-\cos\theta)}}\left[\frac{\ln 4}{\bar\epsilon}+\frac{\bar\epsilon}{12}\right]. 
\ee
The resistivity matrix is obtained using $\boldsymbol{\rho}=\Dv^{-1}\cdot\Iv_{\rm{imp}}\cdot\Dv^{-1}$,
where $\Dv$ is a symmetric $2\times2$ matrix of {\it Drude weights},
which are functions of $\bar\mu$. These functions are
calculated analytically in Ref. \cite{zareniaG} and do not depend on $\tilde v$ since we work within the same Dirac model.  Therefore, the $\tilde v$-dependence of $\rho_{\rm{dis,th}}$ is completely defined through the $\tilde v$-dependence of $\Iv_{\rm{imp}}$,
\be
\rho_{\rm{imp,th}}^{\rm {~long-range}}(\tilde v\to0)\propto\tilde v^2.
\ee
The red curves in Figs. \ref{fig2}(a) and  \ref{fig2}(b), show, respectively, the $\tilde v$- and $T$-dependence of $\rho_{\rm{imp,th}}$ at the CNP ($n=0$). 
We note that $\rho_{\rm{th,dis}}$ can be linked to the disorder electric resistivity $\rho_{\rm{el,dis}}$, through the standard WF relation, i.e. for the Dirac model $\rho_{\rm{el,dis}}/\rho_{\rm{th,dis}}\simeq 2.4 \pi^2k_B^2T/(3e^2)$. While we see that $\Iv_{\rm{imp}}$ in Eq. (\ref{EqIimp})
is independent of $T$, the $T$-dependence of the $\rho_{\rm{imp}}$, see Fig.  \ref{fig2}(b), comes entirely from the $1/T^2$-dependence of the inverse Drude weights.  
\subsubsection{B. Gauge phonons}

In addition to the long-ranged charge impurities, the importance of phonons have been highlighted in the recent theoretical and experimental literature e.g. Refs.~\cite{shaffique2,wu,polshyn}.  We leave a detailed discussion of the differences between the theoretical formulations and the degree to which it explains available experimental data to a forthcoming publication~\cite{girish}.  For our purposes, we need the collision integral for electron-phonon scattering that is given by
\begin{widetext}
\be
\mathbf{I}_{\rm{e-ph}}=\sum\limits _{\mathbf{k}',\lambda'=\pm1}W_{\mathbf{k}'\mathbf{k}}^{\lambda'\lambda}f_{\mathbf{k}',\lambda'}\left(1-f_{\mathbf{k},\lambda}\right)-W_{\mathbf{k}\mathbf{k}'}^{\lambda\lambda'}f_{\mathbf{k},\lambda}\left(1-f_{\mathbf{k}',\lambda'}\right)
\ee
where $W_{\mathbf{k}'\mathbf{k}}^{\lambda'\lambda}$ is the probability
of scattering from state $\ket{\mathbf{k},\lambda}$ to state $\ket{\mathbf{k}',\lambda'}$ (with $\lambda(\lambda')=\pm1$ are band indices),
given by
\begin{align}
W_{\mathbf{k},\mathbf{k}+\mathbf{q}}^{\lambda\lambda'}=\frac{2\pi}{\hbar}\left|g_{\mathbf{k}\mathbf{q}}^{\lambda\lambda'}\right|^{2}\left[N_{\mathbf{q}}\delta\left(\epsilon_{\mathbf{k}+\mathbf{q},\lambda'}-\epsilon_{\mathbf{k},\lambda}-\hbar v_{A}q\right)\right. 
\left.+\left(1+N_{\mathbf{q}}\right)\delta\left(\epsilon_{\mathbf{k}+\mathbf{q},\lambda'}^{m}-\epsilon_{\mathbf{k},\lambda}+\hbar v_{A}q\right)\right],
\end{align}
\end{widetext}
which describes phonon absorption and emission. Here, $N_{\mathbf{q}}=1/\{\exp[\hbar v_{A}q/(k_{B}T)]-1\}$ is the
Bose-Einstein distribution function and $g_{\mathbf{k}\mathbf{q}}^{\lambda\lambda'}$
is electron-phonon coupling, given by
\be
g_{\mathbf{k}\mathbf{q}}^{\lambda\lambda'}=\tilde{\beta}_{A}q\sqrt{\frac{\hbar}{2A\rho\omega_{\mathbf{q}}}}F_{\mathbf{k},\mathbf{k}+\mathbf{q}}^{\lambda\lambda'},
\ee
with effective gauge phonon coupling constant $\tilde{\beta}_A$ and chirality factor $F_{\mathbf{k}\mathbf{k}'}^{\lambda\lambda'}=(1+\lambda\lambda'\cos\theta_{\mathbf{k},\mathbf{k}'})/2$.  Employing the ansatz $f_{\mathbf{k},\lambda}=f_{\mathbf{k},\lambda}^{0}+\lambda eE\tilde{v}\cos\theta_{\mathbf{k}}\tau_{\mathbf{k},\lambda}(\partial f_{\mathbf{k},\lambda}^{0}/\partial\epsilon_{\mathbf{k},\lambda})$,
we obtain (up to linear order) the electron-phonon scattering
time
\be\label{eqtauph}
\frac{1}{\tau_{\rm{e-ph}}^{\lambda}(\mathbf{k})}=\sum\limits _{\mathbf{k}',\lambda'}W_{\mathbf{k}'\mathbf{k}}^{\lambda'\lambda}\frac{1-f_{\mathbf{k}',\lambda'}^{0}}{1-f_{\mathbf{k},\lambda}^{0}}\left(1-\lambda\lambda\cos\theta_{\mathbf{k},\mathbf{k}'}\right),
\ee
which can be evaluated numerically.  For this work we use $\mu_s=7.66\times 10^{-7}$ kg/m$^2$ for the mass density of graphene, and $v_A=1.62\times 10^4$ m/s is the effective acoustic phonon velocity and  $\tilde{\beta}_{A} \approx \beta_{A}\left(\tilde{v} / v_{F}\right) /[2 \tan (\theta / 2)]$ is the effective electron-phonon coupling constant~\cite{shaffique2} with $\beta_A=3.6$ eV is the best estimate for monolayer graphene determined from density functional perturbation theory and tight-binding calculations~\cite{bernevig, mauri}.

The contribution of gauge phonon limited resisitivity  is obtained by averaging
\be
\frac{1}{\rho_{\rm{e-ph}}}=e^{2}\int_{-\infty}^{\infty}d\epsilon N_{D}(\epsilon)\frac{\tilde{v}^{2}}{2}\tau_{\rm{e-ph}}(\epsilon)\left[-f_{0}^{\prime}(\epsilon)\right],
\ee
where $N_D(\epsilon)=4|\epsilon|/[\pi(\hbar\tilde{v})^{2}]$ is the density of states and $\tau_{\rm{e-ph}}(\epsilon)$ is the electron-phonon scattering time (\ref{eqtauph}). At $\tilde{v}/v_A\gtrsim 2$, we can use quasielastic approximations and obtain
\be
\begin{split}
\rho_{\rm{e-ph}}(T\ll T_{\mathbf{BG}})&\simeq 
\frac{1}{\tilde v^2}~\frac{48\zeta(4)\tilde\beta_{A}^{2}(k_{B}T)^{4}}{e^{2}\hbar^{4}\mu_{s}v_{A}^{5} (\pi|n|)^{3/2}},\\
 \rho_{\rm{e-ph}}(T\gg T_{\mathbf{BG}})&\simeq 
\frac{1}{\tilde v^2}~\frac{\pi\tilde\beta_{A}^{2}k_{B}T}{e^{2}\hbar\mu_{s}v_{A}^{2}} .
\end{split}
\ee
In both cases $\rho_{\rm{e-ph}}$ is proportional to $(\tilde{\beta}_{A}/\tilde v)^2\propto 1/\tan^2 (\theta/2)\sim (1-\tilde v)/(1+2\tilde v)$ at fixed $T$ and linearly increases with $T$ for $T\gg T_\mathrm{BG}$, where $T_\mathrm{BG}$ is the Bloch-Gruneisen temperature ($k_B T_\mathrm{BG}=2\hbar v_A k_F$) (see the blue curves in Fig. \ref{fig2}). 
%
%
\begin{figure}[H]
\centering
\includegraphics[width=9. cm]{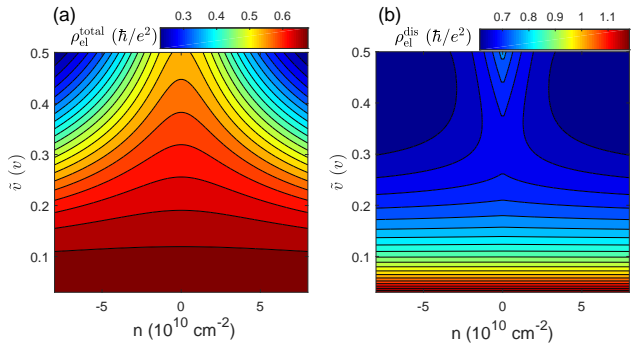}
\caption{ (a) Contour plot of the total electric resistivity, Eq. (\ref{eqRel}) -- including the contributions of charged impurities, gauge phonons, and Coulomb scattering -- as a function of $\tilde v/v$ and doping density $n$ at fixed $T=50$ K.  The charged impurity density is set to $n_{\rm{imp}}=1\times10^{10}$ cm$^{-2}$.
 (b) Same plot for the non-interacting electric resistivity, including only charged-impurity and gauge-phonon scattering: $\rho_{\rm{el}}^{\rm{dis}}=\rho_{\rm{el,imp}}+\rho_{\rm{el,e-ph}}$.  
Observe how the total resistivity in (a) decreases sharply with increasing doping density, in contrast to $\rho_{\rm{el}}^{\rm{dis}}$  in (b), which is weakly density-dependent.  Thus, a sharp peak of $\rho_{\rm{el}}^{\rm{total}}$  around the charge neutrality point constitutes experimental evidence of strong hydrodynamic transport.}  
\label{fig3}
\end{figure} 

In Figs. \ref{fig3}a, and   \ref{fig3}b we respectively show the total electric resistivity $\rho_{\rm{el}}^{\rm{total}}$, defined in Eq. (\ref{eqRel}),  as well as the (b) non-interacting (only the charged impurity and gauges phonon contributions) electric resistivity $\rho_{\rm{el}}^{\rm{dis}}$ as functions of $\tilde v$ and doping density $n$. The total resistivity $\rho_{\rm{el}}^{\rm{total}}$ follows  the Lorentizan form of Eq. (\ref{eqRel}) as a function of density ($n\propto\mu$). 
When $\mu\to0$, the relevant  contribution is the Coulomb term $\rho_{\rm{C}}$, i.e. $\rho_{\rm{el}}\to\rho_{\rm{C}}$ which increases as $\tilde v\to0$. 
In the absence of Coulomb interactions, we observe the  $\tilde v$- and density-dependence of $\rho_{\rm{el}}^{\rm{dis}}$  is completely dominated by the behavior of the total  contributions of  $\rho_{\rm{el,imp}}+\rho_{\rm{el,e-ph}}$ (see Fig. \ref{fig2}a). 
Accordingly, we observe  a minimum in Fig. \ref{fig3}b at $\tilde v\sim 0.4$.  As a function of density, it is interesting to note that 
$\rho_{\rm{el}}^{\rm{dis}}$ is density-independent at small $\tilde v$ where the gauge phonons dominate.  Although the  WF ratio and $1/\Gamma^2$ are the main parameters for identifying the hydrodynamic regime, we note that $\rho_{\rm{el}}$ has a localized peak around the CNP.  This is the opposite behaviour from the non-interacting contributions, and observing this feature in the experiments would be an immediate indication of strong hydrodynamic transport directly from electrical conductivity measurements. 
\subsubsection{C. Calculation of $\Gamma$}
We define the parameter $1/\Gamma^2$, $\Gamma$ is given in Eq. (\ref{eqG}), as a {\it hydrodynamic} parameter which shows the strength of the  momentum-non-conserving disorder collisions vs  the intrinisc electron-hole momentum-conserving Coulomb collisions:
\be\label{EqGamma2}
\frac{1}{\Gamma^2}\propto \frac{\rho_{\rm{C}}}{\rho_{\rm{th,imp}}+\rho_{\rm{th, ph}}}\equiv \frac{\tau_{\rm{eh}}^{-1}}{\tau_{\rm{imp}}^{-1}+\tau_{\rm{ph}}^{-1}}
\ee
\begin{figure}[H]
\centering
\includegraphics[width=8 cm]{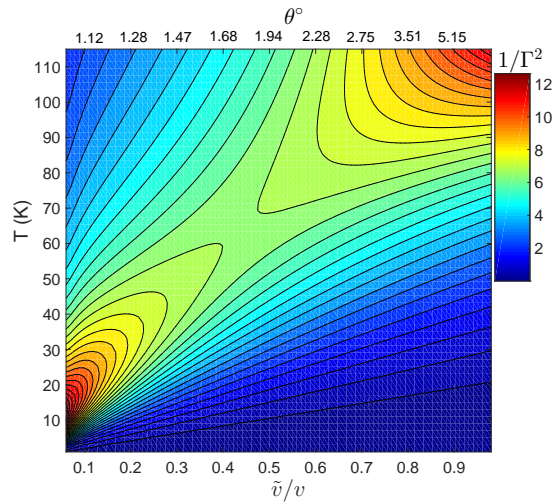}
\caption{Contour plot of  $1/\Gamma^2$ in the ($T-\tilde v$) plane , where $\Gamma$,  defined in Eq. (\ref{eqG}),  determines the width in density of the hydrodynamic transport region.  An enhanced ``hydrodynamic window" characterized by exceptionally large values of $1/\Gamma^2 \simeq 10$ is clearly visible in a range of twist angles on the left side of the figure. 
}
\label{fig4}
\end{figure} 
 Note that at the CNP, $1/\Gamma^2$  is  the physical thermoelectric  parameter $WF(0)$, which is given at finite doping density by the square Lorentzian law \cite{zareniaG}
\be\label{eqWF}
WF(\bar\mu) \simeq \left[\frac{\Gamma}{\Gamma^2+(\alpha\bar\mu)^2}\right]^2 \left(\frac{k_B}{e}\right)^2\,.
\ee
In Fig. \ref{fig4} we mapped out  the $(T-\tilde v)$ 2D-contour plot of $1/\Gamma^2$  at velocities very close to the magic angle and at low temperatures. We point out that $\Gamma$ has a weakly dependence on the doping density $n$ around the CNP (see Fig. \ref{fig3}b). 
When $\tilde v\to0$,  $\rho_{\rm{e-ph}}$ is the dominant disorder mechanism which  linearly increases with $T$. At the other limit $\tilde v\to1$ (which is the case of monolyaer graphene),  $\rho_{\rm{e-imp}}$ becomes important and decays as $1/T^2$ (see Fig. \ref{fig2}). 
Since, $\rho_{\rm{C}}$ is fairly independent of $T$ (has a weakly logarithmic dependency) and slowly varying with $\tilde v$ (see Fig. \ref{fig1}a),  the competition between $\rho_{\rm{e-imp}}$ and  $\rho_{\rm{e-ph}}$ in Eq. (\ref{EqGamma2}), results in the two different regions with stronger hydrodynamic effects ($1/\Gamma^2\gg1$) in Fig. \ref{fig4} : {\it i}) near magic angle and for $10\rm{K} \lesssim T \lesssim 50\rm{K}$ and {\it ii}) $\tilde v\to1$ with $T\gtrsim 70$K (We have neglected the effect of acoustic phonons which are relevant for $T\gtrsim 150$ K in graphene \cite{Morozov} and should suppress the hydrodynamicity as $T$ increases). The crossover between these two regions occurs at $\tilde v\sim 0.4$, associated with the velocity at which $\rho_{\rm{e-ph}}=\rho_{\rm{e-imp}}$ (see Fig. \ref{fig2}a). 
\section{Wiedeman-Franz ratio and Seebeck coefficient}
\begin{figure}
\centering
\includegraphics[width=8.5 cm]{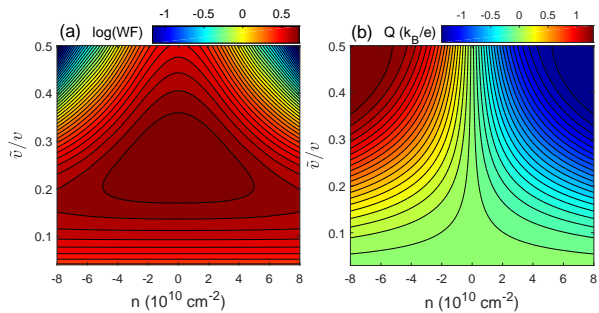}    
\caption{Contour  plot of (a)  $\log$(WF)  and (b) Seebeck coefficient $Q$ as a function of doping density $n$ and Fermi velocity $\tilde v $ at $T=50$ K.  WF is scaled with the Lorentz number $\pi^2 k_B^2/3e^2$ and the charged impurity density is set to $n_{\rm{imp}}=1\times10^{10}$ cm$^{-2}$. The large enhancement of the WF and the peak in the Seebeck coefficient at small $\tilde v$ are both signatures of the hydrodynamic regime.}
\label{fig5}
\end{figure}

 Having all the ingredients we can now calculate the WF, see Eq. (\ref{eqWF}), as well as the Seebeck coefficient, which near the CNP takes the form\cite{zareniaG}
\be\label{eqQ}
Q(\bar\mu)\simeq -\frac{\alpha\bar\mu}{(\alpha\bar\mu)^2+\Gamma^2}\frac{k_B}{e}\,.
\ee
Figures \ref{fig5}a and \ref{fig5}b show respectively the results for the WF and the Seebeck coefficient, including the contribution of both the charged impurity and gauge phonons as functions of doping density $n$ and renormalized Fermi velocity (alias twist angle) $\tilde v$. For clarity  WF is scaled with the Lorentz number $\pi^2 k_B^2/3e^2$ and shown as $\log(\rm{WF})$. 
Consistent with the results in Fig. \ref{fig4}, where at $T=50$K, $1/\Gamma^2$ is large for $0.1\lesssim\tilde v \lesssim 0.45$, we observe a large enhancement of the WF at these near magic twist angle velocities. The broadening of the square-Lorentzian WF peak when $\tilde v\to0$, i.e. approaching the magic twist angle, is caused by the density-independent $\rho_{e-ph}$ which is the dominant disorder scattering mechanism in this regime. 
As seen in Fig. \ref{fig3}b, $\rho_{\rm{dis}}$ becomes  density independent at small $\tilde v$, while it decreases as a function of density as $\tilde v$ increases.  Therefore,  we observe that the Seebeck coefficient in Fig. \ref{fig5}b, which is proportional to $\bar\mu\propto n$, peaks at twist angles corresponding to  $\tilde v\gtrsim 0.4$.  
\section{Concluding remarks}

In this paper, we have calculated the transport properties of twisted bilayer graphene near magic twist angle and at low densities around the charge neutrality point ($K$-point). We have obtained our results for the electric and thermal resistivities,  the Wiedemann-Franz ratio, and the Seebeck coefficient.  Momentum non-conserving scattering mechanisms, such as  long ranged (screened) charge impurities and acoustic gauge phonons, which become most relevant in twisted bilayer graphene near magic angle, are all included in a single parameter $\Gamma \ll 1$, which controls the doping density dependence of the thermoelectric transport coefficients in a region of $\mu/(k_BT)\ll 1$ around the charge neutrality point.  

Our most interesting result is that the hydrodynamic transport anomaly, characterized by large values of the WF ratio and the Seebeck coefficient, is very strong in the vicinity of the magic twist angle and in a temperature range of $10\rm{K} \lesssim T \lesssim 50\rm{K}$, where the gauge phonons are the dominant disorder mechanism.
Furthermore we identify the strong density-dependence of the electric resistivity on a scale controlled by $\Gamma$ near CNP as an unambiguous experimental signature of hydrodynamic transport, noting that no strong density-dependence of the electric resistivity could appear in the absence of dominating electron-hole scattering.\\\\

\begin{acknowledgements}
This work was supported by the U.S. Department of Energy (Office of Science) under grant
No. DE-FG02-05ER46203.  The work in Singapore is
supported by the Singapore Ministry of Education AcRF Tier 2 grants MOE2017-T2-2-140 and MOE2017-T2-1-130, and the National University of Singapore Young Investigator Award (R-607-000-094-133).
\end{acknowledgements}


\end{document}